\documentclass[iop,numberedappendix]{emulateapj}
\slugcomment{Submitted to \apjs} 
\usepackage{natbib}
\usepackage{amsmath}
\usepackage{booktabs}
\usepackage{longtable}
\usepackage{color}
\usepackage{graphicx, subfigure}
\usepackage{hyperref}
\usepackage{CJKutf8}
\usepackage{animate}
\usepackage{graphicx}

\newcommand{\di}{\mathrm{d}} 
\newcommand{\mr}{\mathrm}

\newcommand{\Ho}{\mathrm{H}} 
\newcommand{\Ht}{\mathrm{H_2}} 
\begin{document}
\title{Implementation of chemistry in the Athena++ code}
\author{Munan Gong (\begin{CJK*}{UTF8}{gbsn}龚慕南\end{CJK*})\altaffilmark{1},
Ka-Wai Ho\altaffilmark{2},
James M. Stone\altaffilmark{3}, 
Eve C. Ostriker\altaffilmark{4},
Paola Caselli\altaffilmark{1},
Tommaso Grassi\altaffilmark{1},
Chang-Goo Kim\altaffilmark{4},
Jeong-Gyu Kim\altaffilmark{5},
Goni Halevi\altaffilmark{4}
}

\altaffiltext{1}{Max-Planck Institute for Extraterrestrial Physics,
Garching by Munich, 85748, Germany; 
munan@mpe.mpg.de}
\altaffiltext{2}{Department of Astronomy, University of Wisconsin-Madison, USA}
\altaffiltext{3}{School of Natural Sciences, Institute for Advanced Study, Princeton, NJ 08544, USA}
\altaffiltext{4}{Department of Astrophysical Sciences, Princeton University, Princeton, NJ 08544, USA}
\altaffiltext{5}{Division of Science, National Astronomical Observatory of Japan, Mitaka, Tokyo 181-0015, Japan}

\begin{abstract}
Chemistry plays a key role in many aspects of astrophysical fluids. Atoms and molecules are agents for heating and cooling, determine the ionization fraction, serve as observational tracers, and build the molecular foundation of life. We present the implementation of a chemistry module in the publicly available magneto-hydrodynamic code {\sl Athena++}. We implement several chemical networks and heating and cooling processes suitable for simulating the interstellar medium (ISM). A general chemical network framework in the KIDA format is also included, allowing the user to easily implement their own chemistry. Radiation transfer and cosmic-ray ionization are coupled with chemistry and solved with the simple six-ray approximation. The chemical and thermal processes are evolved as a system of coupled ODEs with an implicit solver from the CVODE library. We perform and present a series of tests to ensure the numerical accuracy and convergence of the code. Many tests combine chemistry with gas dynamics, including comparisons with analytic solutions, 1D problems of the photo-dissociation regions and shocks, and realistic 3D simulations of the turbulent ISM. We release the code with the new public version of { \sl Athena++}, aiming to provide a robust and flexible code for the astrochemical simulation community.
\end{abstract}

\section{Introduction}
Chemistry plays an important role in astrophysical fluids, especially in the ISM, where a diverse range of molecules arise alongside the formation of stars and planets. The chemical species in the ISM act as agents for heating and cooling processes, serve as observational tracers for the physical conditions, control the ionization fraction, and many of them, especially the complex organic molecules, are considered to be related to the origin of life. 

Because of this importance, many numerical tools are developed to simulate chemistry with hydrodynamics (HD) or magneto-hydrodynamics (MHD) in astrophysical fluids. To name a few examples, \citet{Glover2010} modeled chemistry in turbulent molecular clouds using the {\sl ZEUS-MP} code \citep{Norman2000, Hayes2006}; \citet{Walch2015} implemented chemistry in the {\sl FLASH4} \citep{Fryxell2000} code for ISM simulations; \citet{KROME} developed a package to add chemistry to the {\sl ENZO} \citep{Oshea2004}, {\sl RAMSES} \citep{Teyssier2002}, {\sl FLASH} \citep{Fryxell2000}, and {\sl GIZMO} \citep{Hopkins2015, Lupi2021} codes; \citet{Ziegler2016} wrote the chemistry solver in the {\sl NIRVANA} code; \citet{GRACKLE} developed the GRACKLE package for primordial chemistry and cooling; \citet{WangDai2018} and \citet{Wang2019} added chemistry to their own private version of the {\sl Athena++} code \citep{Stone2020} to model protoplanetary disks and planet atmospheres; \citet{Hu2021} studied chemistry of the ISM at different metallicities using the {\sl GIZMO} \citep{Hopkins2015} code; and \citet{Kim2023_photochemistry} developed a photochemistry module for the TIGRESS ISM simulations in the {\sl Athena} code \citep{Stone2008}. It has also become increasingly accessible to model chemistry in (M)HD simulations. Large databases for astrochemistry, such as KIDA \citep{KIDA2010} and UMIST12 \citep{UMIST12}, are publicly available and well-maintained. The growth in computational power and resources in recent years has made expensive chemistry calculations more tractable.

However, there is still a lack of publicly available codes that combine chemistry with (M)HD simulations. Most of the work mentioned above is based on private versions of chemistry codes and/or tailored to specific applications. KROME and GRACKLE are the only public astrochemistry codes, but they are plug-in packages that need to be separately installed and maintained. This is in contrast to the large pool of open-source MHD codes available. In most cases, in order to do chemistry calculations with MHD, it is still up to the individual user to implement their own chemical network and solver from scratch.

In this paper, we present the implementation of an open-source chemistry module in the publicly available MHD code {\sl Athena++} \citep{Stone2020}. {\sl Athena++} is already used for a wide range of applications in astrophysical fluids. It has the advantage of being fast and scalable, providing a flexible framework, and maintaining a public GitHub repository that is continuously updated. {\sl Athena++} also includes many advanced features, such as the adaptive mesh refinement functionality, the general relativity solver, and a GPU version \citep{Grete2021, Grete2022}. Our chemistry module is based on a private version of the code used in \citet{Gong2018, Gong2020_XCO} with significant updates to fully couple chemistry with gas dynamics and provide a code structure suitable for public use. This paper and our accompanying code release aim to offer the community a robust and flexible software package for astrochemistry simulations.

The paper is structured as follows. Section \ref{section:physical_framework} describes the physical framework the code is based on. Section \ref{section:numerical_methods} details the numerical methods we use. Section \ref{section:tests} presents a series of tests performed with our code. Finally, Section \ref{section:summary} summarizes the paper. The public release of our chemistry module is available on the {\sl Athena++} website and GitHub repository (\url{https://www.athena-astro.app/}, \url{https://github.com/PrincetonUniversity/athena}). 

\section{Physical Framework}\label{section:physical_framework}
\subsection{General Equations}
The system of MHD and chemistry equations we solve is, 

\begin{equation}
  \frac{\partial \rho}{\partial t} + \nabla \cdot (\rho \mathbf{v}) = 0,
\end{equation}
\begin{equation}
  \frac{\partial (\rho\mathbf{v})}{\partial t} + \nabla \cdot \left[ \rho\mathbf{v}\mathbf{v} + P^* \mathbf{I} - \frac{\mathbf{B}\mathbf{B}}{4\pi} \right] = 0, 
\end{equation}
\begin{equation}\label{eq:energy}
  \frac{\partial E}{\partial t} + \nabla \cdot \left[ \left( E + P^* \right) \mathbf{v} - \frac{\mathbf{B}(\mathbf{B}\cdot \mathbf{v})}{4\pi} \right] = n(\Gamma - \Lambda),
\end{equation}
\begin{equation}
\frac{\partial \mathbf{B}}{\partial t} - \nabla \times \left( \mathbf{v} \times \mathbf{B} \right) = 0,
\end{equation}
\begin{equation}\label{eq:chemistry}
  \frac{\partial n_s}{\partial t} + \nabla \cdot (n_s \mathbf{v}) = n\mathcal{C}_s.
\end{equation}

Here $\rho$ is the gas density, $\mathbf{v}$ is the gas velocity, $\mathbf{B}$ is the magnetic field, $P^* = P + B^2/(8\pi)$ is the sum of the gas pressure and the magnetic pressure, 
$E = e + \rho v^2/2 + B^2/(8\pi)$ is the total gas energy density, and $e$ is the internal energy density.

In the energy conservation Equation \eqref{eq:energy}, $n=\rho / (\mu_\Ho m_\Ho)$ is the number density of hydrogen nuclei, where $\mu_\Ho$ is the mean molecular weight per hydrogen nucleus, and $m_\Ho$ is the mass of the hydrogen atom. We adopt a constant $\mu_\Ho=1.4$ considering helium and ignoring the contribution from metals. $\Gamma$ and $\Lambda$ are the heating and cooling rates per hydrogen nucleus, respectively. \footnote{We follow the notation in \citet{GOW2017} for $\Gamma$ and $\Lambda$, which also corresponds to the heating and cooling functions implemented in {\sl Athena++}. In \citet{Kim2023_photochemistry}, the notation of $\Lambda$ is different by a factor of $n$.} The heating and cooling processes are detailed in Section \ref{section:heating_cooling}. We adopt an equation of state for the ideal gas with $e=P/(\gamma-1)$, using a constant $\gamma=5/3$.\footnote{We ignore the rotational and vibrational degrees of freedom in $\Ht$. In reality, these additional degrees of freedom are excited at temperatures of $T\gtrsim 80~\mathrm{K}$, leading to changes in $\gamma$ depending on the gas temperature and the $\Ht$ ortho-to-para ratio in warm molecular gas \citep{Boley2007}.}

In the chemistry Equation \eqref{eq:chemistry}, $n_s = n x_s$ is the number density of species $s$, and $x_s$ is the species concentration relative to the hydrogen nuclei. 
All species are assumed to be well-coupled by collisions. For example, there is no drift velocity between ions and neutrals.
The chemical reaction source term $\mathcal{C}_s$ is described in Section \ref{section:chemical_network}. Gas temperature is related to pressure by $P=\rho k_B T/(\mu m_\Ho)$, where $\mu$ is the mean molecular weight per gas particle. Considering the conservation equation of hydrogen atoms, $x_\Ho + x_\mathrm{H^+} + 2 x_\Ht$ = 1, a total helium abundance of $x_\mathrm{He,tot}=0.1$ and an electron abundance of $x_e$, we have $\mu = \mu_\Ho / (1.1 + x_e - x_\Ht)$.

\subsection{Chemical networks\label{section:chemical_network}}
\begin{table*}[htbp]
    \centering
    \caption{Chemical Networks}
    \label{table:network}
    \begin{tabular}{cc cc}
        \tableline
        \tableline
        Name &Number of Species &Reference &Notes\\
        \tableline
        H2 &2 &Section \ref{section:H2_test} &$\Ho$-$\Ht$ network with analytic solution\\
        GOW17 &18 &\citet{GOW2017} 
        &Carbon and oxygen network for the ISM\\
        G14SOD   &9 &\citet{KROME}
        &Primodial chemistry for the modified shock tube test\\
        KIDA &user specified &\citet{KIDA2010}
        &Flexible network in the KIDA format\\
        \tableline
        \tableline
    \end{tabular}
\end{table*}

In general, the chemical reaction source term can be written as,
\begin{equation}\label{eq:dxsdt}
\frac{\di x_s}{\di t} = \mathcal{C}_s(n, T, x_s, Z_d, Z_g, \chi, \xi_\Ho),
\end{equation}
where $Z_d$ and $Z_g$ are the dust and gas metallicity relative to the solar neighborhood, respectively, $\chi$ is the far-ultraviolet (FUV) radiation field strength in \citet{Draine1978} units, and $\xi_\Ho$ is the primary cosmic-ray ionization rate per H nucleus.

The set of chemical species $x_s$ and reaction rates $\mathcal{C}_s$ is defined in the chemical network. Table \ref{table:network} lists the chemical networks implemented in our code and used in the tests (see Section \ref{section:tests}) of this paper. In addition, we provide an interface for users to implement their own chemical network. Below we briefly describe the currently implemented networks, and refer the reader to the respective reference for detailed descriptions.
\begin{itemize}
    \item The H2 network is a simple 2-species reaction network for conversions between H and $\Ht$. Although this network is very simplified and does not represent the realistic H-$\Ht$ transition in the ISM, it has the advantage of having an analytic solution. Thus, it is very useful for testing numerical errors and the convergence of the chemistry solver.
    \item The GOW17 network is originally described in \citet{GOW2017}. It is a widely used and well-tested network for carbon and oxygen chemistry in the atomic and molecular ISM. It has the advantage of being relatively simple with only 18 species and about 50 reactions compared to more sophisticated networks, while still accurately capturing the most important chemical and thermal processes. The GOW17 network is used in many tests in Section \ref{section:tests}.
    \item The G14SOD network is a primordial chemistry network originally implemented in \citet{KROME} with 9 species and 20 reactions. We use this network for the modified shock tube test in Section \ref{section:sod_test}.
    \item The KIDA network allows the user to construct flexible chemical networks simply by providing text files describing chemical species and reactions in the KIDA format \citep{KIDA2010}. We provide an example of the GOW17 network written in the KIDA format. This general KIDA network has the advantage of being adaptable and easy to use. The user can look up chemical reactions from the KIDA website (\url{https://kida.astrochem-tools.org/}) and construct a network without modifying the code. It has the disadvantage of being slower than the hard-coded chemical networks, since the code structure is optimized for flexibility instead of efficiency.
\end{itemize}

\subsection{Heating and cooling processes\label{section:heating_cooling}}
The heating and cooling rates per hydrogen nucleus, $\Gamma$ and $\Lambda$, depend on the chemical and physical properties of the gas, 
\begin{equation}\label{eq:Gamma}
\Gamma = \Gamma(n, T, x_s, Z_d, Z_g, \chi, \xi_\Ho),
\end{equation}
\begin{equation}\label{eq:Lambda}
\Lambda = \Lambda(n, T, x_s, Z_d, Z_g, \chi, |\di v/\di r|),
\end{equation}
where $|\di v/\di r|$ is the velocity gradient used in evaluating the radiative trapping effect for CO cooling \citep{GOW2017}.

The heating and cooling processes implemented vary in different chemical networks. In the H2 network, a simple dust cooling function is included, providing an analytic solution for temperature evolution (see Section \ref{section:H2_test}). In the G14SOD network, molecular hydrogen cooling is included \citep{GA2008, Glover2015}. In the GOW17 network, the major heating and cooling processes for the atomic and molecular ISM are implemented to capture realistic thermodynamics in simulations, including heating from cosmic-rays, photoelectric effect on dust grains, $\Ht$ UV pumping, formation, and photodissociation; and cooling from $\mr{C^+}$, C and O fine structure lines, Lyman-$\alpha$, CO rotational lines, $\Ht$ rovibrational lines and collisional dissociation, H collisional ionization, and recombination of electrons on polycyclic aromatic hydrocarbons (PAHs) \citep[see][and references therein]{GOW2017}. The flexible KIDA network has a default heating and cooling function that is the same as the GOW17 network.

\subsection{Radiation and Cosmic-rays\label{section:radiation}}
Radiation fields and cosmic-rays affect chemistry and heating/cooling in the ISM. In the GOW17 network, the far-ultraviolet (FUV) radiation leads to photoionization and photodissociation of atoms (C and Si) and molecules (CO, $\Ht$, $\mr{CH_x}$ and $\mr{OH_x}$), the UV pumping for $\Ht$, and the photo-electric effect on dust. Cosmic-rays provide a source of ionization and heating. We implement these processes as part of the GOW17 network following the original prescription of \citet{GOW2017}.

In dense gas, FUV radiation is attenuated by dust and molecules. The low-energy cosmic-rays, which dominate the gas ionization in regions shielded from the FUV radiation, lose their energy by interacting with the gas. We calculate both the attenuation of FUV radiation and the cosmic-ray ionization rate using the six-ray radiation transfer method following the implementation in \citet{Gong2018, Gong2020_XCO}, which is further described in Section \ref{section:six-ray}. 

\section{Numerical Methods}\label{section:numerical_methods}

\subsection{Chemistry and Internal Energy}
We solve the source terms for chemical reactions (Equation \eqref{eq:dxsdt}) and the energy source term $\di e/\di t = n(\Gamma-\Lambda)$ (Equations \eqref{eq:energy}, \eqref{eq:Gamma}, and \eqref{eq:Lambda}) as a set of coupled ordinary differential equations (ODEs) in the operator splitting manner. The chemical abundances $x_s$ and the internal energy $e$ are updated after the advection of the chemical species as passive scalars. Two ODE solvers are implemented in the code:
\begin{itemize}
    \item The CVODE solver is an open-source external library developed by \citet{CVODE} (see also \url{https://computing.llnl.gov/projects/sundials/cvode}). It is a general ODE solver using backward differentiation formulas and variable-order, variable-step multistep methods to solve ODEs implicitly. It is especially suited for stiff ODE systems often encountered in chemistry. In order to use this solver, the user needs to install the CVODE library separately from {\sl Athena++}. This is the default ODE solver for the numerical tests in Section \ref{section:tests}.
    \item The simple forward Euler solver uses the explicit first-order differentiation formula to solve the ODEs by sub-cycling. We require the chemical abundances and internal energy to change less than 10\% in each sub-step update. Since the ODE systems are often stiff in chemistry, the forward Euler solver is often unsuitable for realistic problems. We implement the forward Euler solver as a demonstration for the user to write their own ODE solver that is not dependent on the CVODE library. It is only used in the $\Ht$ formation test in Section \ref{section:H2_test} in this paper.
\end{itemize}

The computational cost of solving chemistry using CVODE is dominated by  matrix inversion required by the implicit solver, which scales
roughly as the number of species cubed. Thus, in realistic problems, such as using the GOW17 chemistry in ISM simulations, the chemistry calculation often dominates the total computational cost (see Section \ref{section:turbulent_ism_test}).

We note that, in general, our numerical scheme does not guarantee the exact conservation of elemental abundances and charge. As noted by \citet{Glover2010}, this is mainly because the reconstruction step in the passive scalar advection (left-hand-side of Equation \eqref{eq:chemistry}) is calculated independently for each chemical species. In the GOW17 network, we avoid this problem by using the conservation laws to calculate the abundances of H, He, C, O, Si, and $\mr{e^-}$. Although this simple approach has the advantage of reducing the number of species actively tracked by the ODE solver and thus reducing the computational cost, it has the disadvantage of preferentially concentrating the numerical error into these selected species.
Another solution is to use the consistent multifluid advection (CMA) scheme to re-normalize the flux for passive scalars and ensure the conservation laws are satisfied \citet{PlewaMueller1999, Glover2010}. However, the CMA scheme has the potential disadvantage of introducing additional numerical error during the re-normalization step. We plan to implement and test the CMA scheme in the future.

\subsection{Six-ray Radiation Transfer\label{section:six-ray}}
We use the six-ray radiation transfer method to calculate the shielding of FUV radiation and cosmic-rays. This method is first implemented in \citet{NL1997, NL1999} and \citet{GM2007} and also described in \citet{Gong2018, Gong2020_XCO}. The radiation transfer includes effects from dust shielding, self-shielding of $\Ht$ \citep{DB1996}, self-shielding and cross-shielding from $\Ht$ of CO \citep{Visser2009}, and self-shielding of neutral C \citep{TH1985}. The radiation field is calculated by ray tracing and averaged over the six directions along the Cartesian axes. The incident radiation field is assumed to come from the edge of the computational domain along each ray. The six-ray method has the advantage of being low in computational cost, and \citet{Safranek-Shrader2017} found that the six-ray approximation gives reasonably accurate results for diffuse radiation in ISM simulations when compared to ray tracing along many more different angles. 

We allow the incident radiation field to be separately specified in different directions by the user, and thus the six-ray method can also be used in calculating photo-dissociation region (PDR) structures as shown in Section \ref{section:static_pdr} and \ref{section:moving_pdr}. We update the radiation using the operator splitting method before the chemistry and MHD updates. In the future, we plan to implement more accurate radiation solvers, such as the adaptive ray tracing method described in \citet{Kim2017_ART, Kim2023_photochemistry}.

\section{Tests}\label{section:tests}
In this section, we show results from applying our chemistry module to a series of test problems. Section \ref{section:H2_test} uses a simple 2-species H-$\Ht$ network with an analytic solution to test the numerical accuracy and convergence of the code. Sections \ref{section:pdr_test} and \ref{section:sod_test} show 1D tests of photo-dissociation regions (PDRs) and shock tubes, and compare our results to those from other codes. We show applications of chemistry to realistic 3D simulations of the ISM in Sections \ref{section:tigress_test} and \ref{section:turbulent_ism_test}.

\subsection{$\Ht$ formation\label{section:H2_test}}
\begin{figure}[htbp]
\centering
\includegraphics[width=0.95\linewidth]{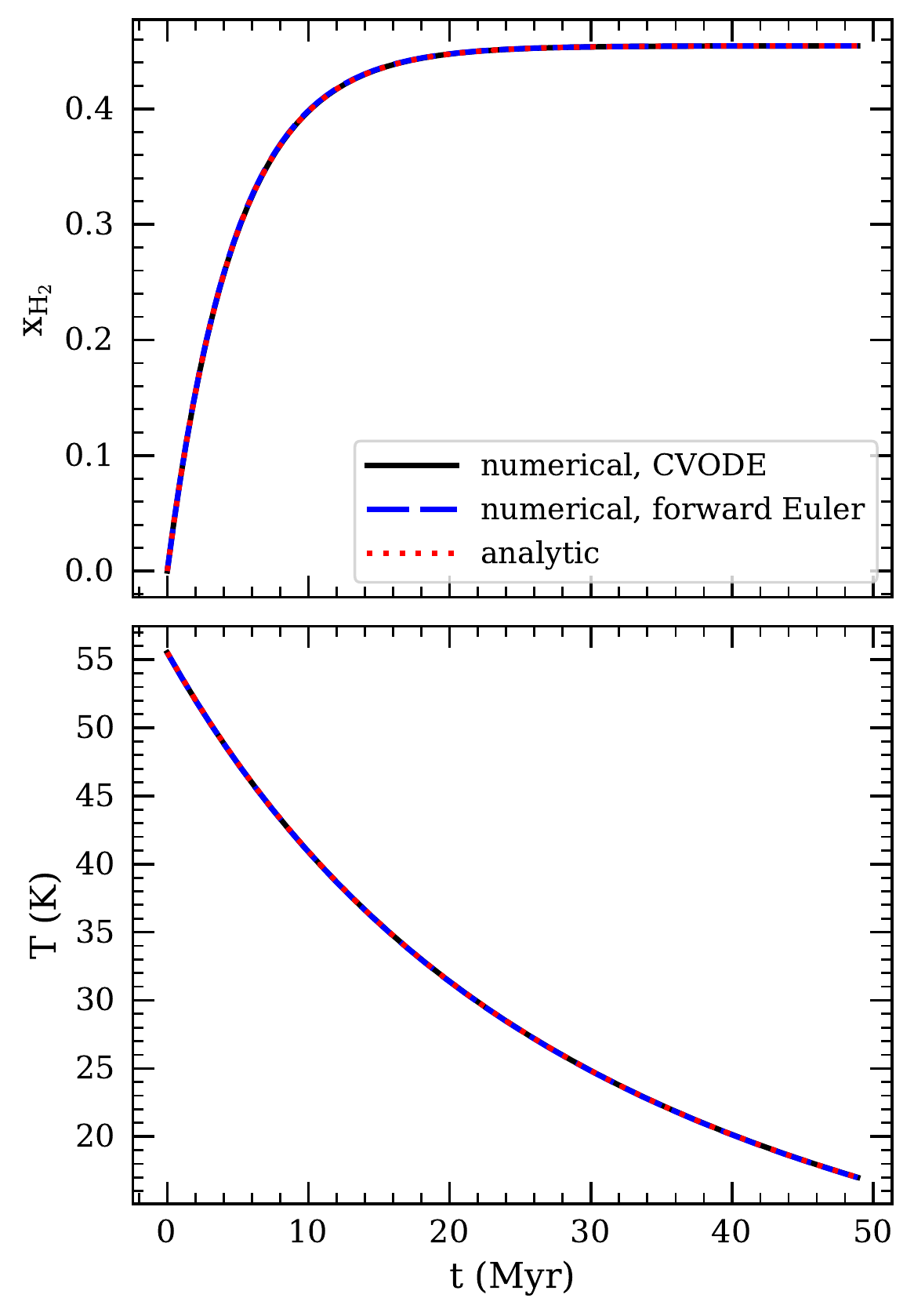}
\caption{Evolution of molecular hydrogen abundance $x_\Ht$ and gas temperature $T$ for the one-zone $\Ht$ formation test. The analytic solutions given by Equations \eqref{eq:fH_t} and \eqref{eq:Tg_t} are shown as the red dotted curves. The black solid and blue dashed curves show the numerical results from the CVODE and forward Euler solvers, respectively. Both solvers yield results that agree with the analytic solutions.}
\label{fig:H2_T_uniform_evolution}
\end{figure}

\begin{figure}[htbp]
\centering
\includegraphics[width=0.95\linewidth]{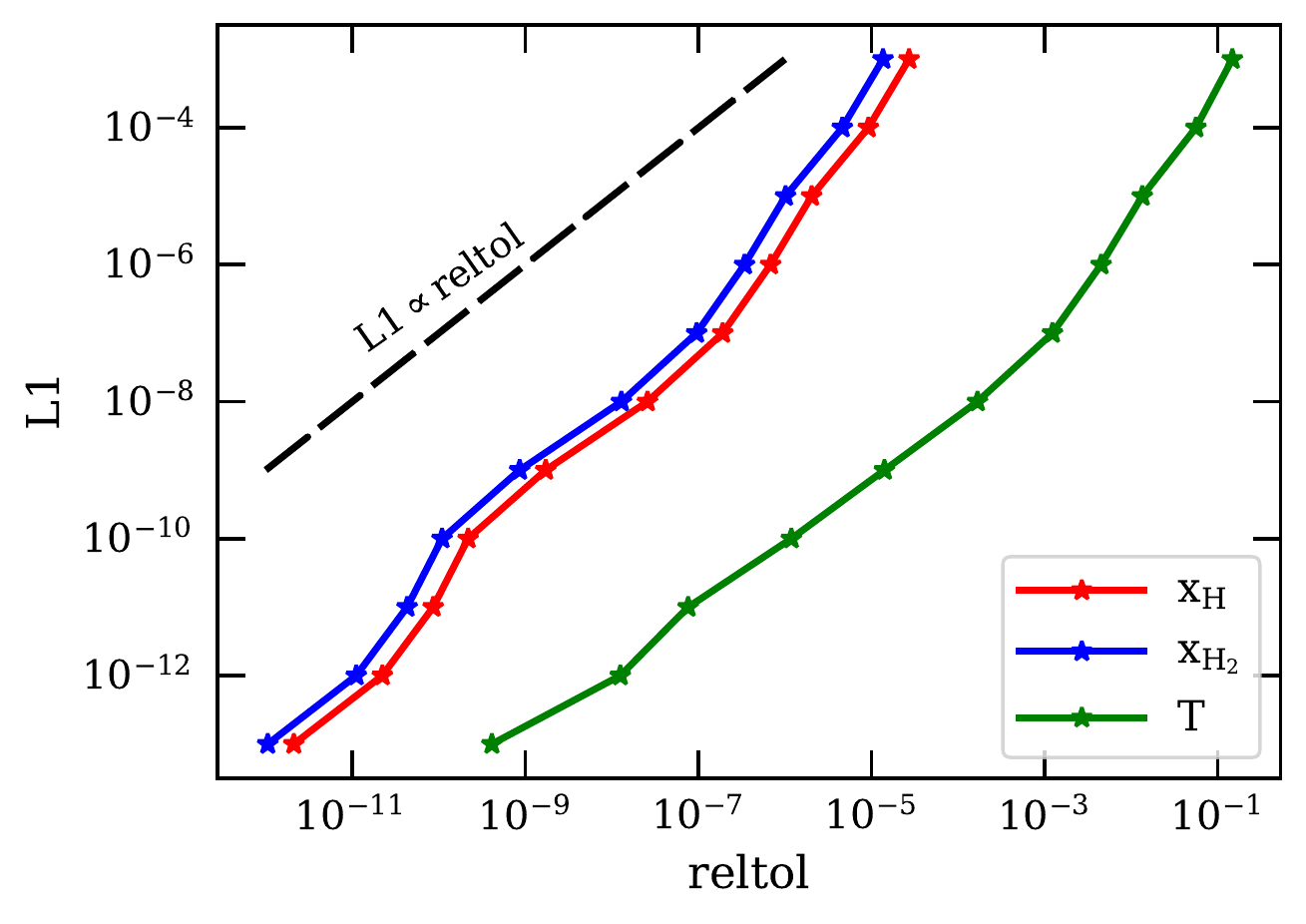}
\caption{L1 error of the numerical solution as a function of the relative tolerance parameter in the CVODE solver for the one-zone $\Ht$ formation test. The gas temperature $T$ is in Kelvin. The L1 error increases when the relative tolerance is less stringent.}
\label{fig:H2_T_uniform_error}
\end{figure}

\begin{figure}[htbp]
\centering
\includegraphics[width=0.95\linewidth]{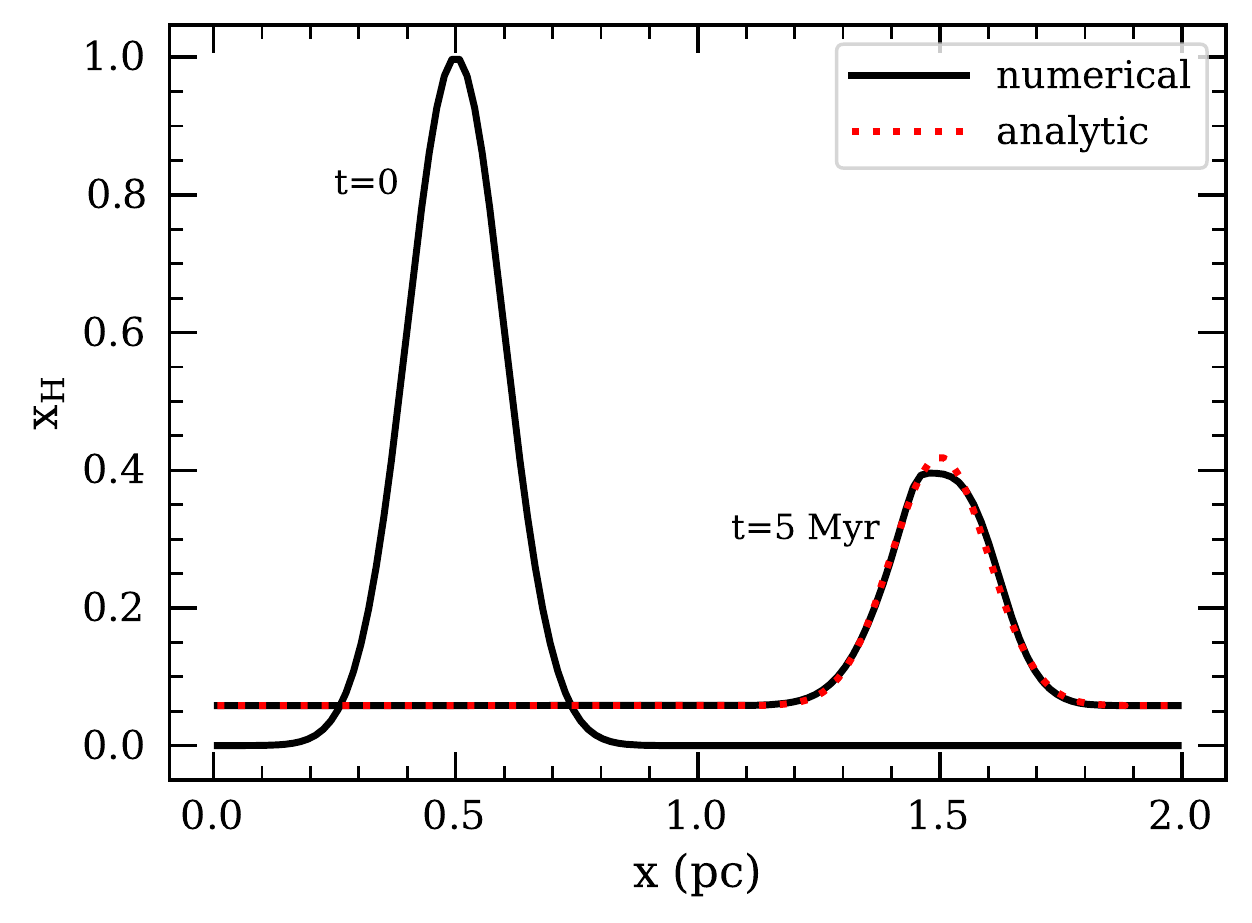}
\caption{Abundance profile of the atomic hydrogen $x_\Ho$ in the $\Ht$ formation with advection test. After 5 Myr, the numerical result is compared with the analytic solution. Here the number of grid cells is $N=128$, and we use the RK2 integrator with PLM reconstruction.}
\label{fig:fH_advection}
\end{figure}

\begin{figure*}[htbp]
\centering
\includegraphics[width=0.95\linewidth]{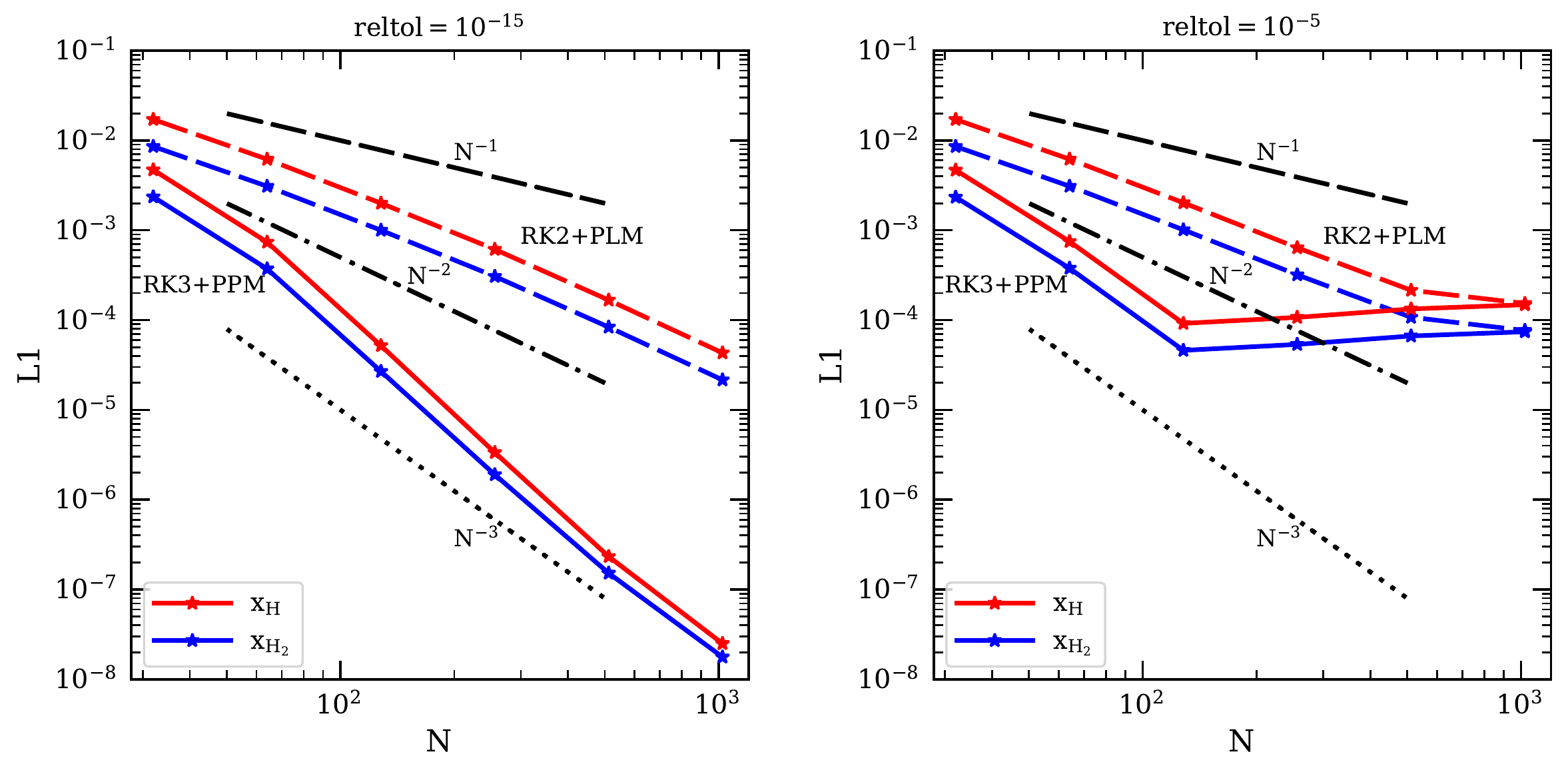}
\caption{Convergence of chemical abundances for the $\Ht$ formation with advection test. The L1 error of $\Ho$ and $\Ht$ abundances are shown as red and blue lines, respectively. The results using the second-order scheme (RK2+PLM) are shown as dashed lines, while results from the third-order scheme (RK3+PPM) are in solid lines. As the resolution increases, the numerical error from advection decreases until, above a threshold resolution, the error from the chemistry ODE solver dominates and the error flattens.}
\label{fig:H2_advection_convergence}
\end{figure*}

In this test, we construct a simple network of $\Ht$ formation that
contains two species $\Ho$ and $\Ht$, and two reactions, the $\Ht$ formation on
dust grains
\begin{equation}\label{eq:H2_gr}
    \mr{H + H + gr \rightarrow H_2 + gr}, 
    ~ k_\mr{gr}=3\times 10^{-17}~\mr{cm^3 s^{-1}},
\end{equation}
and $\Ht$ destruction by cosmic-rays 
\begin{equation}\label{eq:H2_cr_original}
    \mr{H_2 + cr \rightarrow H_2^+ + e},
    ~ k_\mr{cr}=3\xi_\Ho,
\end{equation}
where $\xi_\Ho$ is the primary cosmic-ray ionization rate per H nucleus. \footnote{The rate of Reaction \ref{eq:H2_cr_original} is $k_\mr{cr} = 2 \xi_\Ho (2.3 x_\Ht + 1.5 x_\Ho)$ in \citet{GOW2017}. In order to simplify the analytic solution, we use a constant rate of $k_\mr{cr}=3\xi_\Ho$.}

Reaction \ref{eq:H2_cr_original} is followed by two reactions, 
\begin{align}
    \mr{H_2^+ + H_2} &\rightarrow \mr{H_3^+ + H}, \label{eq:H2plus_H2}\\
    \mr{H_3^+ + e}    &\rightarrow \mr{H_2 + H}.   \label{eq:H3plus_e}
\end{align}
Because Reactions \ref{eq:H2plus_H2} and \ref{eq:H3plus_e} are fast $(k
\gtrsim 10^{-9}~\mr{cm^3~s^{-1}}$ for both), we use the summed reaction
\begin{equation}\label{eq:H2_cr}
    \mr{H_2 + cr \rightarrow H + H}
\end{equation}
to represent Reactions \ref{eq:H2_cr_original}, \ref{eq:H2plus_H2} and
\ref{eq:H3plus_e}, with the rate of Reaction \ref{eq:H2_cr_original}.

From Reactions \ref{eq:H2_gr} and \ref{eq:H2_cr}, we can write down the rate
equation for $x_\Ho$,
\begin{equation}
    \frac{\di x_\Ho}{\di t} = -2x_\Ho n k_\mr{gr} + 2x_\Ht k_\mr{cr}.
\end{equation}
Along with the conservation of hydrogen atom,
\begin{equation}\label{eq:H_cons}
    x_\Ho + 2x_\Ht = 1,
\end{equation}
the evolution of $x_\Ho$ and $x_\Ht$ can be solved analytically:
\begin{equation}\label{eq:fH_t}
    x_\Ho(t) = \left( x_\mr{H,0} - \frac{a_2}{a_1}\right)e^{-a_1 t} 
                  + \frac{a_2}{a_1},
\end{equation}
where $a_1 = k_\mr{cr} + 2nk_\mr{gr}$, $a_2 = k_\mr{cr}$, $x_\mr{H,0}$ is the
$\Ho$ abundance at $t=0$, and 
$x_\Ht = (1-x_\Ho)/2$. At $t=\infty$, the steady state solution gives $x_\Ho = a_2/a_1$. The
timescale to reach the steady state is $1/a_1$.

To test the temperature evolution, we include a simple dust cooling function given by 
\begin{equation}
    \Lambda_\mr{dust} = \alpha_\mr{gd} n T^{1/2}(T - T_d),
\end{equation}
where $\alpha_\mr{gd}$ is the dust-gas coupling coefficient, $T$ is the gas
temperature, and $T_d$ is the dust temperature. In the case of $T \gg T_d$, we have 
$\Lambda_\mr{dust} \approx \alpha_\mr{gd} n T ^{3/2}$. With $T = e_\mathrm{g,sp}/c_{v,\Ho}$,
where $e_\mathrm{g,sp}=e/n$ and $c_{v,\Ho}$ is the specific heat capacity per H nucleus,  we can write
\begin{equation}
    \frac{\di e_\mathrm{g,sp}}{\di t} = - \alpha_\mr{gd} n \left( \frac{e_\mathrm{g,sp}}{c_{v,\Ho}}
    \right)^{3/2}.
\end{equation}
In principle, both $\alpha_\mr{gd}$ and $c_{v,\Ho}$ depend on the $\Ht$ fraction.
However, in order to obtain an analytical solution, we use a constant
$\alpha_\mr{gd}=3.2\times 10^{-34}~\mr{erg~s^{-1}cm^3 K^{-3/2}}$ and $c_{v,\Ho} =
1.65 k_B$ (assuming atomic hydrogen and a helium fraction of 0.1) \citep{GOW2017, Goldsmith2001}. In
this case, the analytic solution for internal energy evolution can be obtained,
\begin{equation}
    e_\mathrm{g,sp}(t) = \left[ \frac{\alpha_\mr{gd} n c_{v,\Ho}^{-3/2}}{2}(t-t_0) + e_0^{-1/2}
    \right]^{-2},
\end{equation}
where $e_0\equiv e_\mathrm{g,sp}(t=0)$.
Equivalently, for the temperature evolution,
\begin{equation}\label{eq:Tg_t}
    T(t) = \left[ \frac{\alpha_\mr{gd}n}{2c_{v,\Ho}}(t-t_0) + T_0^{-1/2}
    \right]^{-2},
\end{equation}
where $T_0 \equiv T(t=0)$.

We first test the chemistry and gas temperature evolution in a one-zone model without hydrodynamics, with the initial condition of $x_\Ho = 1$, $x_\Ht = 0$ and $T = 55.5~\mathrm{K}$. We use a constant cosmic-ray ionization rate of $\xi_\Ho=2\times 10^{-16}~\mr{s^{-1}}$ and density of $n=100~\mr{cm^{-3}}$. The evolution of $\Ht$ abundance and gas temperature is shown in Figure \ref{fig:H2_T_uniform_evolution}. The numerical results from both the CVODE and the forward Euler solvers match the analytic solutions. Since hydrodynamics is not included, the numerical error originates only from the chemistry ODE solver. In CVODE, the numerical error is controlled by two parameters: the relative tolerance ``reltol'' and the absolute tolerance ``abstol''. We vary the relative tolerance values and use a fixed absolute tolerance of $10^{-20}$.\footnote{In general, more relaxed tolerance results in a reduction in the computational cost, with the relative tolerance often having the larger impact on the CVODE solver speed. It is recommended to set a relatively small absolute tolerance, since the solver may be unstable and the abundance of trace species may be incorrect otherwise.}
Figure \ref{fig:H2_T_uniform_error} shows the L1 error of the numerical result at the end of the simulation, defined as,
\begin{equation}
    L1 = \frac{1}{V_\mr{tot}}\sum_i |\Delta f_i|\Delta V_i,
\end{equation}
where $\Delta f = f_\mr{numerical}-f_\mr{analytic}$ is the error between the numerical result and analytic solution for quantity $f$ in each cell, $\Delta V$ is the cell volume, and $V_\mr{tot}=\sum_i \Delta V_i$ is the total volume of the simulation domain.
As the relative tolerance becomes less stringent, the L1 error increases.

To test the interaction between the chemistry ODE solver and the hydrodynamics solver, we set up a 1D advection test with the  $\Ht$ formation chemistry. To ensure a simple analytic solution, we use an isothermal equation of state with a fixed temperature of 55.5 K and a uniform background with a density of $n=100~\mr{cm^{-3}}$, a velocity of $v_x=0.2~\mr{km~s^{-1}}$, and a cosmic-ray ionization rate of $\xi_H = 2\times 10^{-16}~\mr{s^{-1}}$. We use a uniform Cartesian grid along the x-axis. The simulation domain contains $x=[0,2~\mathrm{pc}]$, and the boundary condition is periodic.
The initial $\Ho$ abundance is a Gaussian profile centered around
$x=0.5~\mathrm{pc}$ and with width $\sigma=0.1~\mathrm{pc}$. For simplicity, we set the initial $x_\Ho=0$ in
$x=[1~\mathrm{pc},2~\mathrm{pc}]$. We vary the number of grid cells
$N$ and the order of the advection method to test the numerical convergence. The second-order method uses the second-order Runge–Kutta (RK2) integrator and the piece-wise linear method (PLM) for reconstruction. The third-order method uses the third-order Runge–Kutta (RK3) integrator and the piece-wise parabolic method (PPM) for reconstruction.

We compare the simulation results with the analytic solution at $t=5~\mathrm{Myr}$, as shown in Figure \ref{fig:fH_advection}. The numerical convergence is shown in Figure \ref{fig:H2_advection_convergence}. There are two sources of error: the advection of species abundances, and the chemistry integration by the CVODE solver. The advection error is determined by the choice of hydrodynamic solver and numerical resolution. The tolerance parameters in CVODE control the error from solving the chemistry ODEs. As the numerical resolution increases, the total error decreases at first since it is dominated by the advection error. However, eventually, the total error becomes dominated by the CVODE solver error and therefore stays roughly constant with increased numerical resolution.

\subsection{Photo-dissociation Region\label{section:pdr_test}}
\subsubsection{Static PDR\label{section:static_pdr}}
\begin{figure*}[htbp]
\centering
\includegraphics[width=0.99\linewidth]{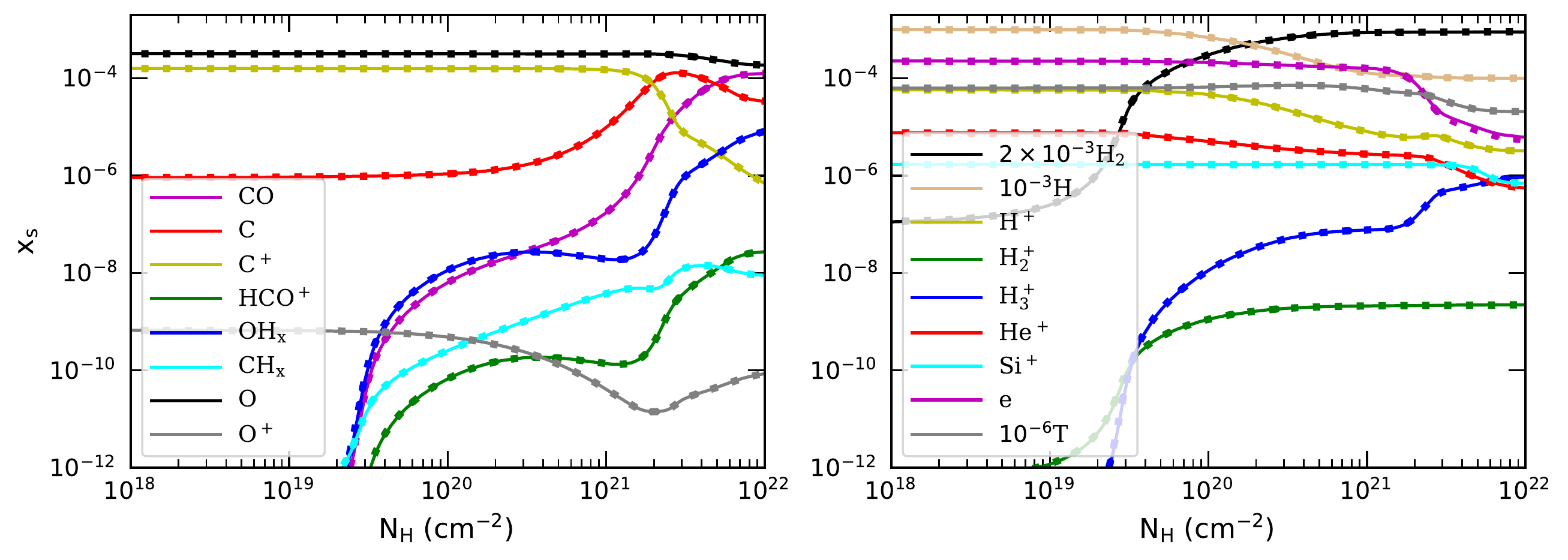}
\caption{Static one-sided PDR test. The abundances of chemical species and gas temperature are shown as lines with different colors indicated by the legends. The thin solid lines show the result from our chemistry code with the GOW17 network. %TODO: refer to chemical network table
The thick dotted lines, which overlap with the thin solid lines, show the results from the open-source PDR code by \citet{GOW2017}.}
\label{fig:pdr_static}
\end{figure*}

Modeling PDRs is one widely used application of ISM chemistry. In our first PDR test, we set up a static 1D one-sided PDR, where the FUV radiation illuminates one side of a uniform slab of ISM. We adopt parameters similar to the conditions of the cold neutral medium in the solar neighborhood: the solar metallicity, a uniform background density of $n=100~\mathrm{cm^{-3}}$, an incident FUV radiation field of $\chi_0=1$ in the \citet{Draine1978} units, and a uniform primary cosmic-ray ionization rate of $\xi_\Ho=2\times 10^{-16}~\mathrm{s^{-1}}$ \citep{Indriolo2007}. The GOW17 network is used to model the hydrogen, carbon, and oxygen chemistry, as well as the gas heating and cooling. The hydrodynamics is turned off, and the chemistry, temperature, and radiation field are evolved for 10 Gyr until the steady state is achieved. We compare our result with that from the open-source PDR code \footnote{https://sites.google.com/view/munangong/codes/pdr-code} developed by \citet{GOW2017}. 

The six-ray radiation module is used to calculate dust and molecular shielding of the radiation field. We set the incident radiation field to $\chi_0=6$ in the +$x$ direction, and zero in all other 5 directions. Since the radiation field strength is an average of the 6 directions in the six-ray method, this is equivalent to an incident radiation field of $\chi_0=1$ perpendicular to the slab surface in the 1D one-sided slab set up in the PDR code by \citep{GOW2017}. We note that a relatively high spatial resolution is needed to resolve the H-to-$\Ht$ transition, which occurs at a column density of $N_\Ho \approx 10^{19}~\mathrm{cm^{-2}}$ \citep{DB1996}. We use a uniform Cartesian grid with $10^4$ cells in the x-direction within $x=[0, 40~\mr{pc}]$ in our chemistry code, covering $N_\Ho = 10^{18} - 10^{22}~\mathrm{cm^{-2}}$. 
In the \citet{GOW2017} PDR code, we use a logarithmically spaced grid between 
$N_\Ho = 10^{17} - 10^{22}~\mathrm{cm^{-2}}$ with $10^5$ cells. We checked that the spatial resolution in both codes is enough to resolve the H-to-$\Ht$ transition and other major chemical and thermal structures. 
The exact setup in the two codes can be found in their respective Git repositories\footnote{See the regression test {\sl chemistry/chem\_pdr\_static} in the {\sl Athena++} code, and file {\sl run/athenapp\_benchmark.cpp} in the \citet{GOW2017} PDR code.}.

The resultant chemical and thermal structure of the PDR is shown in Figure \ref{fig:pdr_static}. As the column density increases, the FUV radiation is shielded by dust and molecules. This causes the abundances of molecular species to increase while the electron abundance and temperature decrease. The chemical abundances and gas temperature in the two codes agree very well with each other. We note that the two different codes have their own advantages for modeling PDRs: {\sl Athena++} allows more flexible configuration of the simulation domain and it is straightforward to model other geometries such as two-sided slabs (see Section \ref{section:moving_pdr}) or PDRs in 2D and 3D. The \citet{GOW2017} PDR code is restricted to and specifically optimized for 1D one-sided PDRs. It is faster and has an easier setup for logarithmically spaced grids in column density which are often more suitable for resolving chemical structures.

\subsubsection{Moving PDR\label{section:moving_pdr}}
\begin{figure*}[htbp]
\centering
\includegraphics[width=0.99\linewidth]{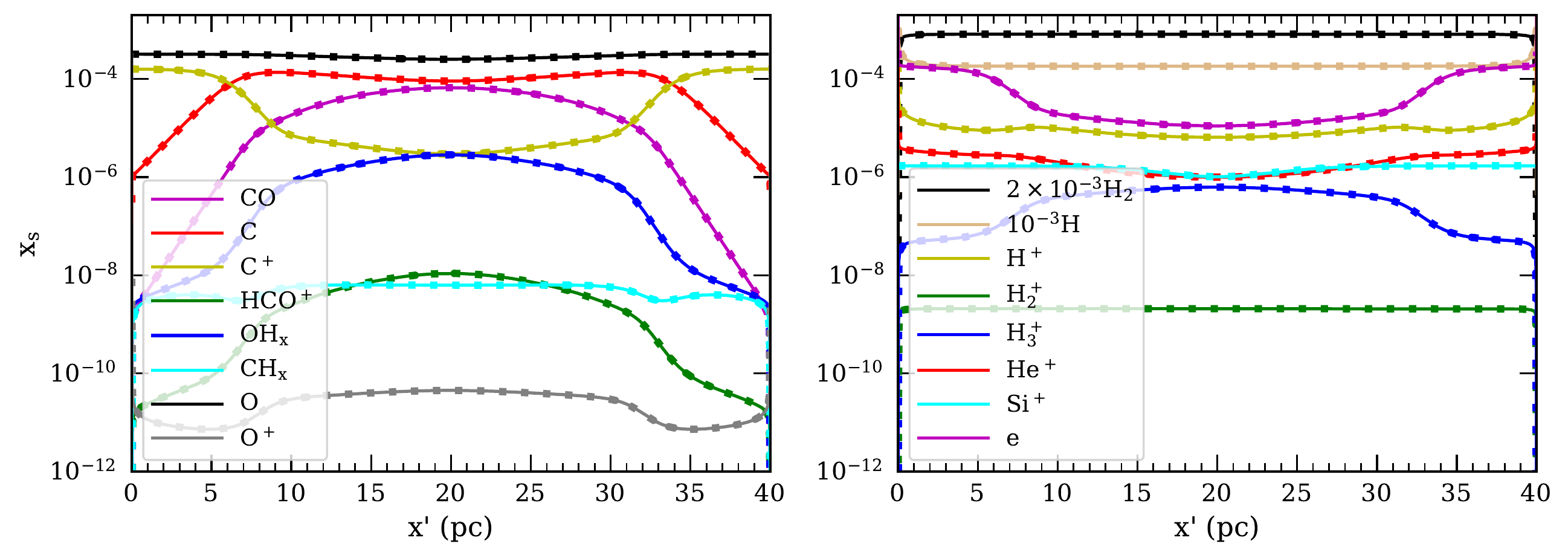}
\caption{Moving 2-sided PDR test. The abundances of chemical species at $t=1~\mathrm{Myr}$ are shown as lines with different colors indicated by the legends. The x-axis is the relative position of the PDR in the cold medium. The thin solid lines show the static reference solution with $v_x=0$. The thick dotted lines, which overlap with the reference solution, show the results from the moving PDR with $v_x = 5~\mathrm{km/s}$.}
\label{fig:pdr_moving}
\end{figure*}

To further test the interaction between the chemistry and the hydrodynamic solvers, we construct a 1D two-sided moving PDR test. Initially, a uniform cold cloud with $n_c=100~\mathrm{cm^{-3}}$ and $T_c=40~\mathrm{K}$ located at $x=[5~\mathrm{pc}, 45~\mathrm{pc}]$ is embedded in the hot surrounding medium. The background medium with $n_w=0.1~\mathrm{cm^{-3}}$ and $T_w=4\times 10^4~\mathrm{K}$ extends outside of the cold cloud within $x=[0, 55~\mathrm{pc}]$. The incident FUV radiation field is set to $\chi_0=6$ in the $+x$ and $-x$ directions in the six-ray radiation module, and zero in all other 4 directions. The warm medium provides negligible shielding of radiation field due to its low density, and the PDR structure forms in the cold cloud. A uniform primary cosmic-ray ionization rate is set to be $\xi_\Ho=2\times 10^{-16}~\mathrm{s^{-1}}$. We use the GOW17 chemical network at the solar metallicity, same as in Section \ref{section:static_pdr}. The initial species concentration is set to $10^{-6}$ for $\mr{H^+}$, $\mr{H_2^+}$, $\mr{H_3^+}$, $\mr{He^+}$, $\mr{O^+}$, $\mr{C^+}$, $\mr{CO}$, $\mr{HCO^+}$, $\mr{Si^+}$, $\mr{CH_x}$ and $\mr{OH_x}$, and 0.4 for $\Ht$. The abundances of $\Ho$, $\mr{e-}$, $\mr{O}$, $\mr{C}$, $\mr{He}$, $\mr{Si}$ are obtained with conservation laws. A uniform Cartesian grid with 13750 cells is used, so that the spatial resolution is the same as in Section \ref{section:static_pdr}, and the H-to-$\Ht$ transition is resolved.

In order to maintain the pressure equilibrium between the cold cloud and the hot surrounding medium, we manually set a constant $\mu=2.33$ and zero heating and cooling rates throughout the simulation. We first obtain a stationary reference solution with the velocity $v_x=0$ and fix the hydrodynamic variables. Then we set a uniform $v_x=5~\mathrm{km/s}$ to model the moving PDR. Both the reference solution and the moving PDR are run for $t=1~\mathrm{Myr}$.

The result for the moving PDR test is shown in Figure \ref{fig:pdr_moving}. The $x$-axis shows the relative position within the cold cloud, $x'=x-x_0-v_x t$, where $x_0=5~\mathrm{pc}$. Since the incident FUV radiation field is symmetric (neglecting the shielding from the warm medium), the distribution of the chemical species is also symmetric, with each side similar to the one-sided PDR in Section \ref{section:static_pdr}. Small differences from the one-sided PDR result from the fact that the chemistry has not quite reached the steady state (although it is close) at 1 Myr, and that the cold cloud is not infinite in size so the radiation field has contributions from both sides within the cloud. The moving PDR solution agrees with the static reference solution as expected.

\subsection{Modified Shock Tube Test\label{section:sod_test}}
\begin{figure}[htbp]
\centering
\includegraphics[width=0.95\linewidth]{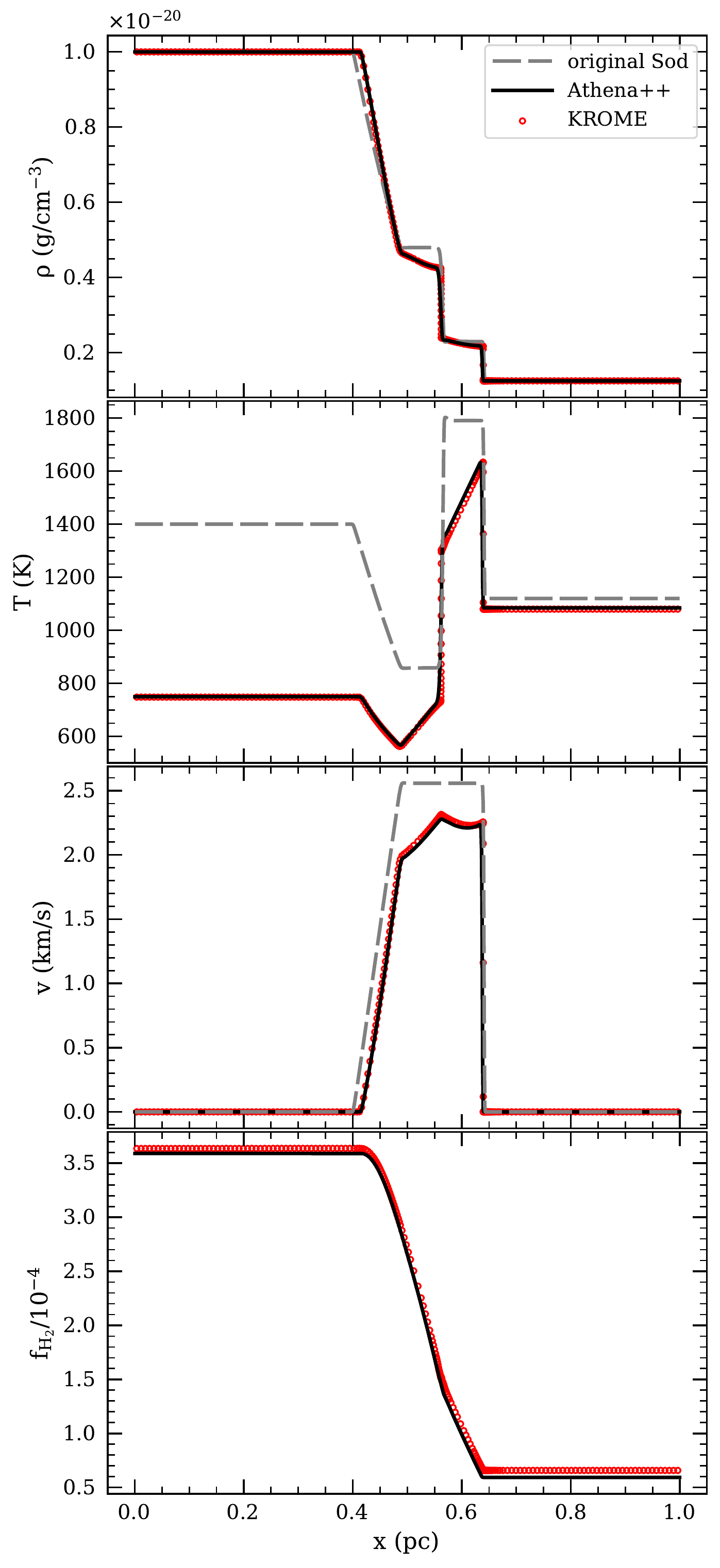}
\caption{Results of the modified shock tube test from our {\sl Athena++} code (black solid curves), the {\sl KROME} code \citet{KROME} (red circles), and the original Sod shock tube without chemistry and cooling (gray dashed curves). From top to bottom, the panels show the gas density, temperature, velocity and the mass fraction of $\Ht$. }
\label{fig:G14AthComparsion}
\end{figure}

\begin{table}[htbp]
\centering
\caption{Initial abundance $x_s$ of chemical species in the modified shock tube test.}
\label{table:G14Sod_init}
\begin{tabular}{lc}
\tableline
\tableline
Species                    & $x_s$      \\
\tableline
H                               & 0.999890 \\
$\mr{H^+}$                      & $1.084475 \times 10^{-4}$\\
$\mr{He}$                       & 0.080605 \\
$\mr{He^+}$                     & 0      \\
$\mr{He^{2+}}$                  & 0      \\
$\mr{H^-}$                      & 0      \\
$\mr{H_2}$                      & $9.99890\times 10^{-7}$\\
$\mr{H^+_2}$                    & 0     \\
$\mr{e^-}$                      & $1.084475 \times 10^{-4}$\\
\tableline
\tableline
\end{tabular}
\end{table}

The modified 1D Riemann problem combines a classical Sod shock test with additional chemistry and cooling. In this test, we adopt the setup in the {\sl euler1D} test in the {\sl KROME} code.\footnote{The source code for the test in {\sl KROME} can be found at \url{https://bitbucket.org/tgrassi/krome/src/master/tests/euler1D} (commit 3169cde).}
We use the primordial chemical network described in \citet{KROME} (see their Table C1) and $\Ht$ rovibrational cooling \citep{GA2008, Glover2015}. The initial condition is set to be two separate regions at rest within the simulation domain of 1 pc in size. The gas in the left-side region has a higher density and pressure, which generates the shock. The initial density and temperature are $\rho(x)=10^{-20}~\mr{g\:cm^{-3}},~T(x) = 1400~\mr{K}$ at $x\leq 0.5~\mr{pc}$; and $\rho(x)=1.25\times 10^{-21}~\mr{g\:cm^{-3}},~T(x) = 1120~\mr{K}$ at $x > 0.5~\mr{pc}$.
The initial abundance $x_s$ of each species $s$ is listed in Table \ref{table:G14Sod_init}. The mean molecular weight is kept at a constant value of 1.25 following the treatment of this specific test in {\sl KROME}.
In both {\sl Athena++} and {\sl KROME}, the simulation is performed using the RK2 time integrator and the second-order PLM reconstruction scheme in uniform Cartesian grids, with a numerical resolution of $N = 1024$. The HLLC Riemann solver is used in {\sl Athena++}, and the more diffusive HLL solver is used in {\sl KROME}.

Figure \ref{fig:G14AthComparsion} shows the results from the simulation at $t=2.45 \times 10^4~\mathrm{yr}$, and the comparison with the results from {\sl KROME}. In contrast to the original Sod tube test, the additional cooling results in lower temperatures and different shock profiles. Because $\Ht$ is the coolant, the differences are larger where the $\Ht$ abundance is higher. Our results from {\sl Athena++} agree well with those from \citet{KROME}. Since there is no analytic solution for this problem, and the ODE solver, Riemann solver, and implementation of $\Ht$ cooling is slightly different between the codes, it is difficult to make an exact comparison between the two codes. The source code for this test in {\sl Athena++} is available as the regression test {\sl chemistry/chem\_G14Sod}.

%A convergence test is also performed to test the result between the resolution. We pick resolution $N=8192$ as a reference point and compare the result with different resolution ranging from 64 to 4096 with factor of 2 between the resolution. Fig. \ref{fig:convergence_sod} showed the result. We compute the L2 error of $H_2$ mass fraction to test the convergence. We define it as :$\Delta \varepsilon^2 = <(X-X_{ref})^2>$. A linear decline trend is observed with increasing the resolution throughout, showing the solution is converging with higher resolution.

%Ka-Wai Ho, comparison with \citet{Ziegler2016} and \citet{KROME}

\subsection{Post-processing TIGRESS simulations\label{section:tigress_test}}
\begin{figure*}[htbp]
\centering
\includegraphics[width=0.99\linewidth]{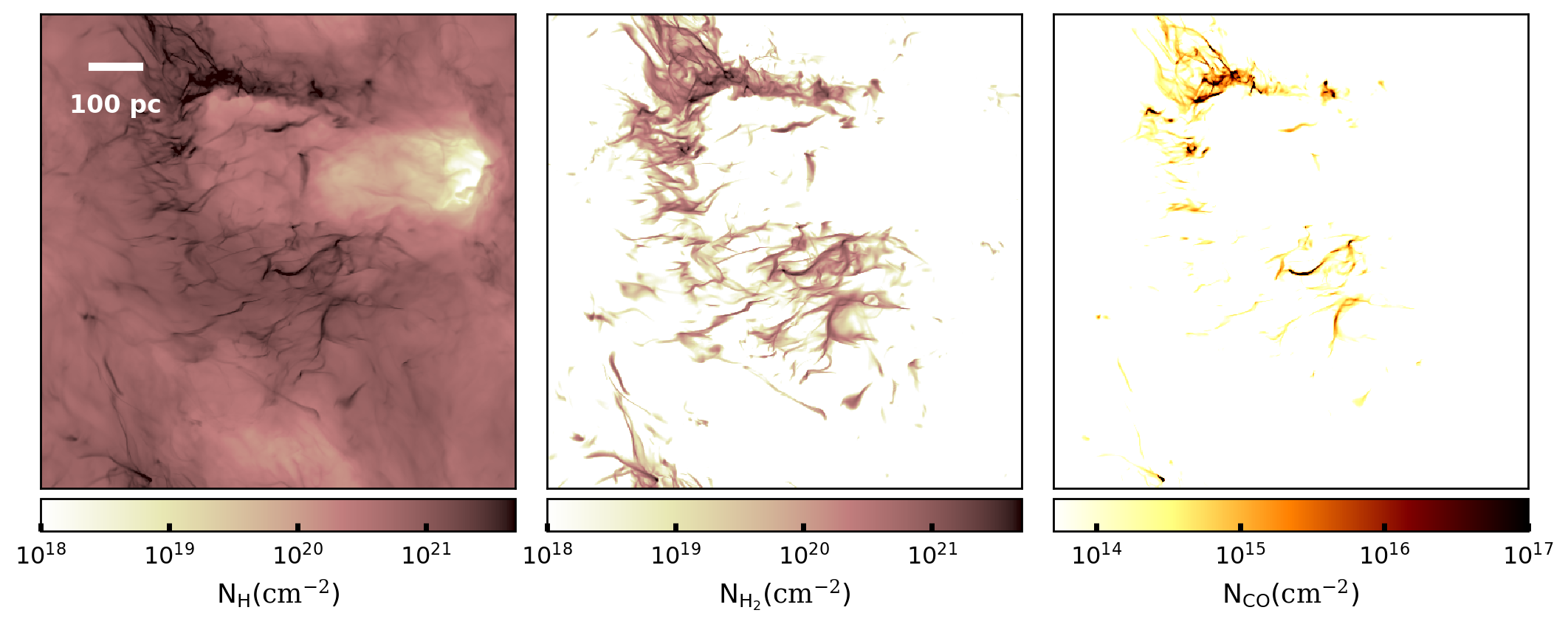}
\caption{One snapshot from TIGRESS simulations of the solar neighborhood ISM and molecular content from chemistry post-processing. The left panel shows the column density from the original TIGRESS output. The middle and right panels show the $\Ht$ and CO column densities from chemistry post-processing.}
\label{fig:ism_tigress}
\end{figure*}

One practical application of chemistry is to post-process ISM simulations and obtain molecular abundances. In \citet{Gong2018} and \citet{Gong2020_XCO}, an earlier, but very similar, version of this chemistry module is used to post-process 
%the TIGRESS multi-phase ISM simulations \citep{KO2017} to 
the star-forming ISM simulations using the TIGRESS framework \citep{KO2017}, in which star clusters
form in the magnetized ISM via gas cooling and gravity. Stellar feedback sources turbulence via supernovae and gas heating via FUV photoelectric effect.
The post-processed multiphase ISM results were used to
study the CO to $\Ht$ conversion factor ($X_\mr{CO}$) and its dependence on the galactic environments. Here we demonstrate this capability with our chemistry module.

We obtain the initial density and velocity fields from the TIGRESS simulations of the solar-neighborhood ISM and hold them fixed during the chemistry post-processing. We use the GOW17 network to calculate the chemical and thermal evolution (see Section \ref{section:chemical_network}) and the six-ray radiation transfer method to obtain the shielding for the FUV radiation field and cosmic-rays (see Section \ref{section:radiation}). The uniform incident FUV radiation field and cosmic-ray ionization rate from the edge of the simulation domain is scaled with the star formation rate from the simulation and normalized by the solar neighborhood values \citep[for more details see][]{Gong2018, Gong2020_XCO}. The chemistry and radiation are run iteratively until the steady state is reached. We include the input parameters and configure options for running the code in the file \textit{inputs/chemistry/athinput.chem\_tigress}.

The result from one snapshot of the TIGRESS simulation is shown in Figure \ref{fig:ism_tigress}. Molecular species, such as $\Ht$ and CO, form in the denser part of the ISM where the radiation field is attenuated by dust and molecular shielding. 

Recently, the original TIGRESS simulation and the post-processed chemistry outputs from \citet{Gong2018, Gong2020_XCO} (similar to that shown in Figure \ref{fig:ism_tigress}) have been made publicly available on the TIGRESS data release platform \citep{tigress_data_release}. In addition, the TIGRESS data release includes the post-processed CO line intensities from \citet{Gong2018, Gong2020_XCO} and post-processed ionized gas distribution from \citet{Kado-Fong2020}. We encourage the reader to explore the data release products if ISM chemistry is of interest.

\subsection{Turbulent Multi-phase ISM\label{section:turbulent_ism_test}}
\begin{figure}[htbp]
\centering
\includegraphics[width=0.95\linewidth]{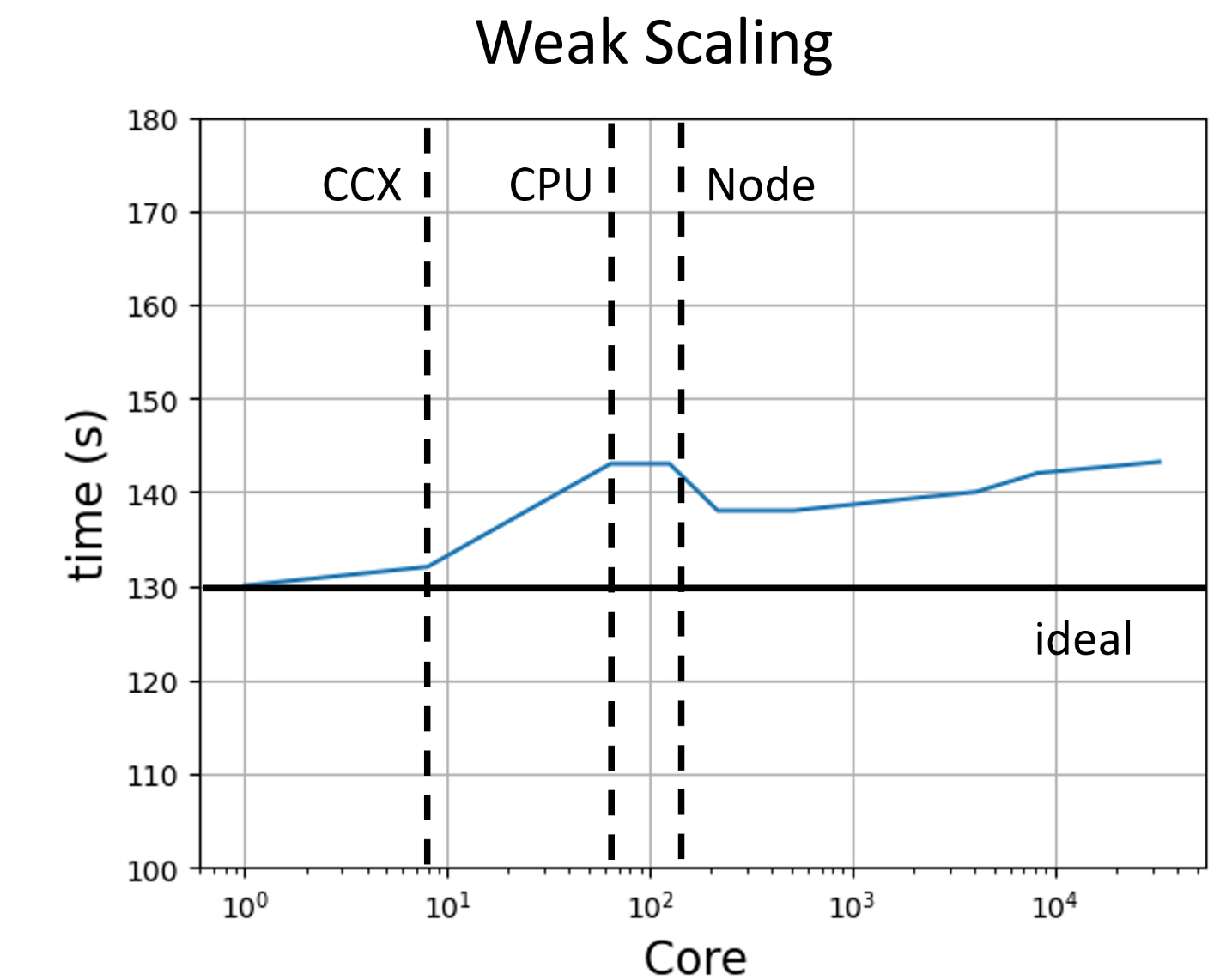}
\caption{Weak scaling of the computational cost for the turbulent multi-phase ISM with time-dependent chemistry test performed using the NERSC Perlmutter cluster. The $x$- and $y$-axis denote the number of cores used and the time for finishing 100 iterations with a fixed block size of $48^3$ for each core, respectively. Each dashed line represents a different potential hardware bottleneck for parallel efficiency, which are the core complex (labeled as CCX, 8 cores for zen3), the CPU package (labeled as CPU, 64 cores in our test), and the compute node (labeled as Node, 128 cores in our test). The black horizontal line is the time for a single CPU core to finish 100 iterations, which is also the reference of ideal parallel efficiency.}
\label{fig:weak_scaling}
\end{figure}

\begin{figure*}[htbp]
\centering
\includegraphics[width=0.99\linewidth]{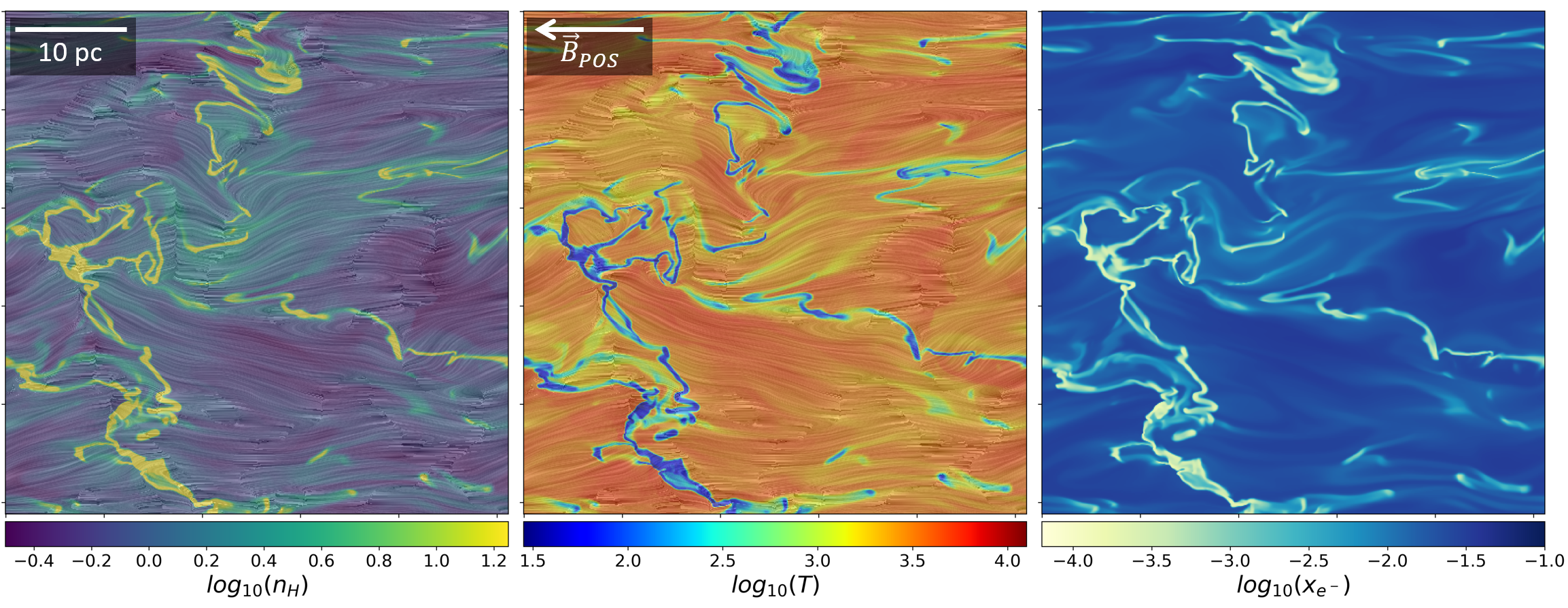}
\caption{Results from the turbulent multi-phase ISM simulation test. The panels show a 2D slice of the gas density (left) and temperature (middle) overlaid with the plane-of-the-sky magnetic field streamline, and the electron abundance (right). The white arrow in the middle panel represents the mean magnetic field direction.}
\label{fig:MP2D}
\end{figure*}

To demonstrate the applicability of our chemistry module in realistic 3D MHD simulations, we present a simulation representing the solar neighborhood ISM. This multi-phase ISM simulation couples the gas dynamics with chemical evolution, as well as cooling and heating effects. %In previous numerical studies,  specific chemical species fractions are assumed and a synthetic cooling function is introduced to manage the cooling and heating effects \citep{KI2002}.

We use a similar initial setup as the weak magnetic field case presented in \citet{10.1093/mnras/stad481}, but with a smaller box size (50 pc) with tri-periodic boundary and slightly different parameters (initial density $n_\mr{H,0} = 2~\mr{cm^{-3}}$, and magnetic field $B_0 = 1~\mu\mr{G}$). Turbulence is driven continuously in the simulation domain with a stochastic forcing method described by \citet{Schmidt2009}. Instead of using a simple cooling function as in \citet{10.1093/mnras/stad481}, we use the GOW17 chemical network (see Section \ref{section:chemical_network}) to model the evolution of chemical species and gas heating and cooling. We use a constant FUV radiation field strength of $\chi=1$ and cosmic-ray ionization rate of $\xi_\Ho=2\times 10^{-16}~\mr{s^{-1}}$ both spatially and temporally.
We run the simulation for 16 Myr, equivalent to about one sound crossing time for the warm neutral medium, to allow the system to reach a steady state.
%We find that our simulation is much more computationally expensive compared to synthetic cooling functions. 
We use a numerical resolution of $512^3$, and the simulation takes 20 hours to finish using 32768 AMD zen3 cores in the NERSC Perlmutter cluster. We note that the additional real-time chemistry calculation with the GOW17 network (18 species, about 50 reactions) is about 5 times more expensive compared to using a simple ISM cooling function in \citet{10.1093/mnras/stad481}. In addition, we present the weak scaling for the computational cost in Figure \ref{fig:weak_scaling}.

Figure \ref{fig:MP2D} shows a 2D slice of the gas density, temperature, magnetic field direction, and electron abundance at the end of the simulation. We observe the formation of two stable phases and one transitional phase: the warm (WNM), cold (CNM), and thermally unstable neutral medium. %The cold neutral medium generally has a low ionization rate of $10^{-4}$, while the warm phase rapidly increases to about $10^{-1}$
%The physical properties of the ISM in our simulations are consistent with expectations from the literature \citep{1998ApJ...494L..19D,2019ApJ...878..157X}. A comparison between the simulated and expected values is presented in Table \ref{table: ISMcomparsion}. 
\begin{table}[htbp]
\centering
\caption{Average density, temperature, and electron abundance of the WNM and CNM in the turbulent multiphase ISM test
%comparison of two phases between the theoretical idealized expectation of ISM from \cite{1998ApJ...494L..19D} (labeled as DL98 
%in the table) and our result. We take the mean of each phase as the value presented in the table.
}
\label{table:ISMcomparsion}
\begin{tabular}{llll}
\tableline
\tableline
                & $n~(\mr{cm^{-3}})$ & $T~(\mr{K})$   & $x_\mr{e}$                     \\ \tableline
%DL98 & 0.4/ 30        & 6000/100 & 0.1/ $10^{-4}$ \footnote{While \citep{1998ApJ...494L..19D} did not provide the explicit value of $x_\mr{e}$, \cite{2019ApJ...878..157X} reported this property and their value is used in this study.}                \\ \hline
 WNM              & 0.7      & 6568 & 0.0258\\
CNM              & 37.8      & 127 & $5.5 \times 10^{-4} $\\ 
\tableline
\tableline
\end{tabular}
\end{table}

Table \ref{table:ISMcomparsion} shows the spatially averaged density, temperature, and electron abundance of the WNM and CNM in the simulation snapshot shown in Figure \ref{fig:MP2D}. The WNM and CNM are defined using temperature cuts at $>5500~\mr{K}$ and $<200~\mr{K}$, respectively. These results are similar to the typical estimates in the solar neighborhood environment from other theoretical and observational literature \citep[e.g.][]{Draine2011, Wolfire1995, HT2003}. A more detailed analysis of our simulations will be presented in a separate paper in the future. 
This test demonstrates the ability of our chemistry module to handle strong non-equilibrium conditions arising from the presence of intense turbulence motion.

\section{Summary}\label{section:summary}
In this paper, we present the implementation of chemistry in the publicly available MHD code {\sl Athena++}. This is one of few open-source astrochemistry codes coupled with fluid dynamics, providing a flexible and robust code suitable for a wide range of applications. 

We describe the physical framework, including equations solved, chemical networks, heating and cooling processes, radiation, and cosmic-rays in Section \ref{section:physical_framework}. Several chemical networks are implemented (Table \ref{table:network}), including the GOW17 network practical for realistic ISM simulations, and the general KIDA network which allows the user to easily implement their own chemistry of choice by simply providing input text files. The numerical methods are described in Section \ref{section:numerical_methods}. We use the operator-split method to solve chemistry and radiation. The source terms for the chemical reactions and heating/cooling are solved as a system of coupled ODEs. We implemented the CVODE solver suitable for stiff ODEs often encountered in chemistry, as well as a simple forward Euler solver ODE solver mainly as a demonstration for the user to write their own ODE solver that does not depend on the external CVODE library. Radiation transfer is calculated with the simple six-ray method where ray tracing is performed along the Cartesian coordinates.

A series of tests for our code is presented in Section \ref{section:tests}. The numerical accuracy and convergence of the code are shown in detail using a 2-species H-$\Ht$ network with an analytic solution. 1D tests of the PDR structure and the modified shock tube problem are performed, and the results are compared with other available chemistry codes. Finally, we show examples of including chemistry in realistic 3D ISM simulations, for both post-processing chemistry and time-dependent chemistry coupled with MHD. 

In the future, we plan to extend our chemistry module with other existing and upcoming features of {\sl Athena++}. For example, an extension of {\sl Athena++} for general equation of state has been made by \citet{Coleman2020} and can be coupled with chemistry to solve problems where the adiabatic index depends on the chemical composition, e.g. for warm molecular gas \citep{Boley2007}.
A more comprehensive cosmic-ray transport method is already implemented in {\sl Athena++} by \citet{JiangOh2018} and \citet{Armillotta2021, Armillotta2022}, which can also be coupled with chemistry. The adaptive ray tracing method for FUV and ionizing radiation has been implemented in the {\sl Athena} code by \citet{Kim2017_ART} and extended with additional chemistry and thermal processes in \citet{Kim2023_photochemistry}, and we plan to make similar functionalities available in {\sl Athena++}.

\section{Acknowledgments}
M. G., P. C., and T. G. acknowledge the financial support of the Max Planck Society. M. G. acknowledges Interstellar Institute's program ``The Grand Cascade'' and the Paris-Saclay University's Institut Pascal for hosting discussions that nourished the development of the ideas behind this work. K.-W. H.  acknowledges the National Energy Research Scientific Computing Center (NERSC) for providing free allocation time during the charging holiday, operated under Contract No. DE-AC02-05CH11231 using NERSC award FES-ERCAPm4239.

\bibliographystyle{apj}
\bibliography{apj-jour, all}
\end{document}